\documentstyle[12pt,epsf]{article}

\if@twoside
\else \oddsidemargin 00pt \evensidemargin 00pt
\fi
\makeatletter
\expandafter\ifx\csname mathrm\endcsname\relax\def\mathrm#1{{\rm #1}}\fi
\@ifundefined{mathrm}{\def\mathrm#1{{\rm #1}}}{\relax}
\makeatother
 
\def\mathswitch#1{\relax\ifmmode{{#1}}\else${#1}$\fi}
\def\mathswitchr#1{\relax\ifmmode{\mathrm{#1}}\else$\mathrm{#1}$\fi}

\newcommand{\Pe}{\mathswitch {e}}
\newcommand{\Pem}{\mathswitch {e^-}}
\newcommand{\Pep}{\mathswitch {e^+}}
\newcommand{\Pf}{\mathswitch f}
\newcommand{\Pt}{\mathswitch t}
\newcommand{\Ptb}{\mathswitch {\bar{t}}}
\newcommand{\PH}{\mathswitch H}
\newcommand{\PZ}{\mathswitch Z}
\newcommand{\PX}{\mathswitch X}
\newcommand{\ttb}{\Pt\Ptb}
\newcommand{\eett}{\mathswitch{\Pep\Pem\to\Pt\Ptb}}
\newcommand{\eettH}{\mathswitch{\Pep\Pem\to\Pt\Ptb\PH}}
\newcommand{\eettZ}{\mathswitch{\Pep\Pem\to\Pt\Ptb\PZ}}
\newcommand{\eettX}{\mathswitch{\Pep\Pem\to\Pt\Ptb\PX}}
\newcommand{\AFB}{\mathswitch {A_{FB}}}
\newcommand{\GeV}{\unskip\,\mathrm{GeV}}
\newcommand{\TeV}{\unskip\,\mathrm{TeV}}

\newcommand{\beq}{\begin{equation}}
\newcommand{\eeq}{\end{equation}}
\newcommand{\bea}{\begin{eqnarray}}
\newcommand{\eea}{\end{eqnarray}}
\newcommand{\nn}{\nonumber}
\def\reffi#1{\mbox{Fig.~\ref{#1}}}
\def\reffis#1{\mbox{Figs.~\ref{#1}}}
\newcommand{\real}{\mbox{Re}}
\newcommand{\al}{\alpha}
\newcommand{\Gmu}{G_{\mu}}

\newcommand{\sw}{s_w}

\newcommand{\mz}{M_\PZ}
\newcommand{\mh}{M_\PH}
\newcommand{\mf}{m_\Pf}
\newcommand{\mt}{m_\Pt}
\newcommand{\pe}{l_1}
\newcommand{\pp}{l_2}
\newcommand{\pt}{p_1}
\newcommand{\ptb}{p_2}
\newcommand{\rob}{\bar{\rho}}
\newcommand{\Szr}{\hat{\Sigma}^Z}
\newcommand{\Sgzr}{\hat{\Sigma}^{\gamma Z}}
\newcommand{\Pigr}{\hat{\Pi}^{\gamma}}

\newcommand{\Pigzr}{\hat{\Pi}^{\gamma Z}}
\newcommand{\Pizr}{\hat{\Pi}^Z}
\newcommand{\fgf}{F^{\gamma f}}
\newcommand{\fzf}{F^{Z f}}

\begin{document}

\begin{title}
\date{}
\title{
\hfill {\small KA-TP-6-1996} \\[1cm]
Top pair production in  $e^+e^-$ collisions with virtual \\
and real electroweak radiative corrections.%
\thanks{To appear in the Proceedings of the Workshop "Physics
with $e^+e^-$ Colliders", Annecy-Gran Sasso-Hamburg 1995, ed.~P.~Zerwas}
}
\author{V. Driesen, W. Hollik, A. Kraft \\[0.2cm] 
\and
Institut f\"ur Theoretische Physik\\
Universit\"at Karlsruhe, D-76128 Karlsruhe, Germany}
\maketitle
\date{}
\begin{abstract}
The effect of virtual electroweak corrections to $e^+e^- \rightarrow 
t \bar{t}$ and the  contribution of the radiation processes 
$e^+e^- \rightarrow t\bar{t}Z, t\bar{t}H$ to the inclusive top 
pair production cross section and forward-backward asymmetry are 
discussed in the high energy regime.
\end{abstract}
\end{title}
\thispagestyle{empty}
\clearpage
\setcounter{page}{1}

\vskip 1cm
\begin{center}
{\Large \bf \boldmath Top pair production in \Pep\Pem\ collisions
with virtual and real electroweak radiative corrections. \par}
\vskip 2.5em
{\large
{\sc V.~Driesen, W.~Hollik, A.~Kraft } \\[1ex]
{\it Institut f\"ur theoretische Physik \\
     Universit\"at Karlsruhe \\
     D-76128 Karlsruhe, Germany} \par}
\vskip 1em
\end{center}

{\abstract 
The effect of virtual electroweak corrections to $e^+e^- \rightarrow 
t \bar{t}$ and the  contribution of the radiation processes 
$e^+e^- \rightarrow t\bar{t}Z, t\bar{t}H$ to the inclusive top 
pair production cross section and forward-backward asymmetry are 
discussed in the high energy regime. } 

\begin{figure}[ht]
\begin{center}
\mbox{
\begin{tabular}{ccc}
\epsfxsize5cm
\epsffile[140 310 500 480]{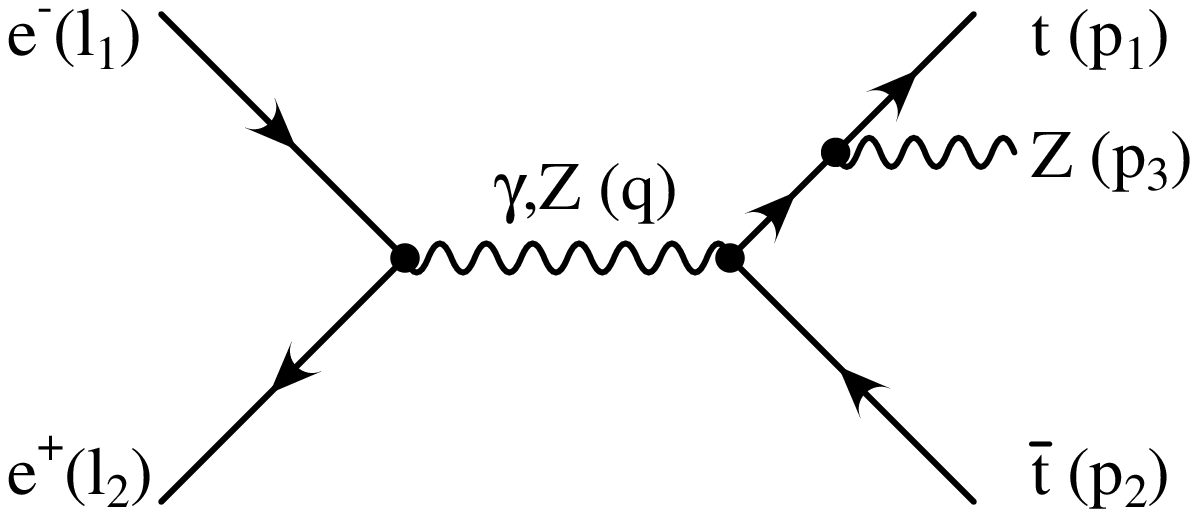}
\epsfxsize5cm
\epsffile[140 310 500 480]{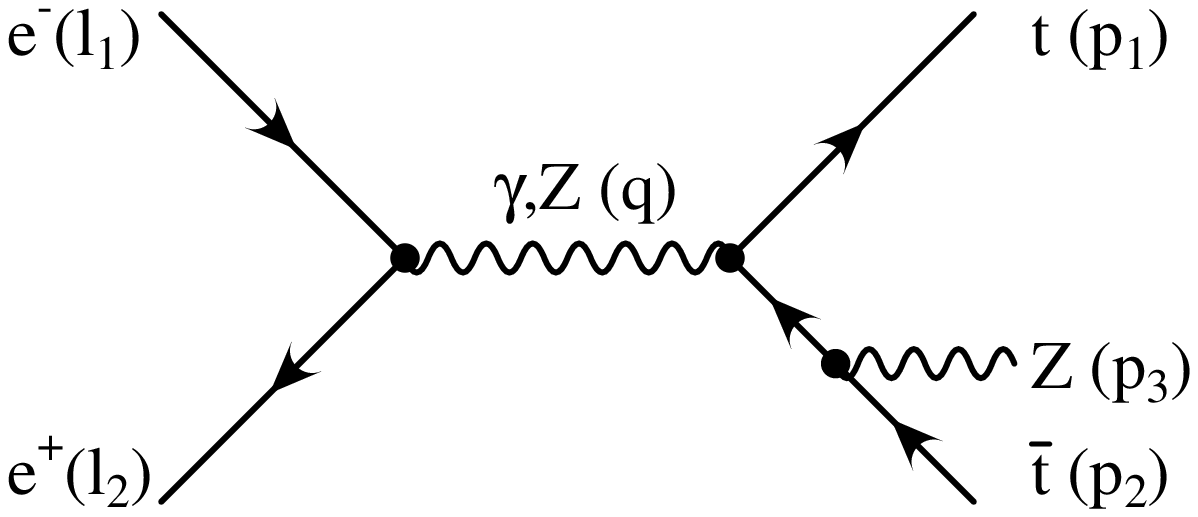}
\epsfxsize5cm
\epsffile[140 310 500 480]{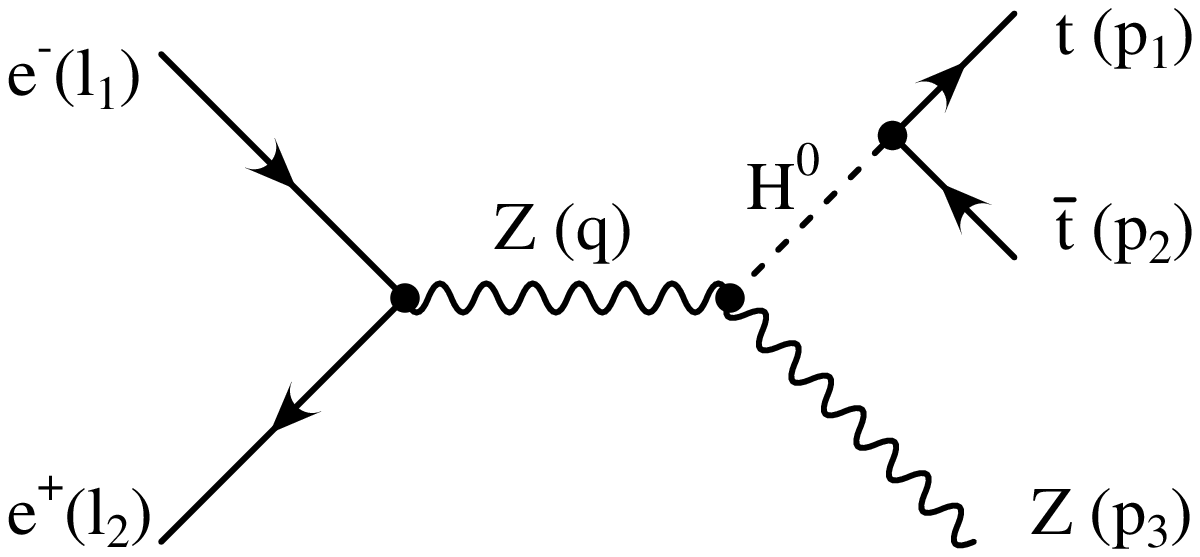}
\end{tabular}
}
\mbox{
\begin{tabular}{cc}
\epsfxsize5cm
\epsffile[140 310 500 480]{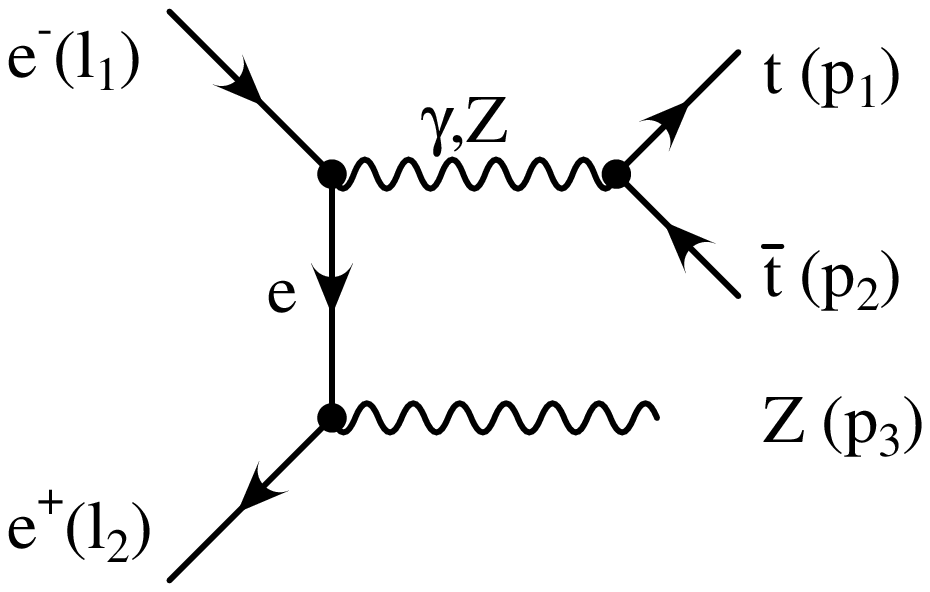}
\epsfxsize5cm
\epsffile[140 310 500 480]{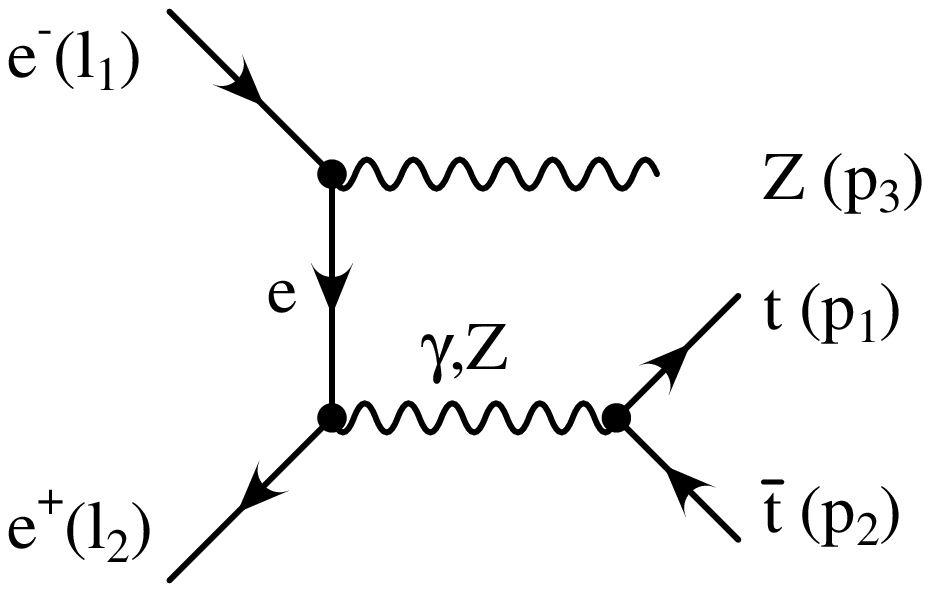}
\end{tabular}
}
\caption{\label{figz}\it Diagrams for \eettZ.}
\end{center}
\end{figure}

\begin{figure}[ht]
\begin{center}
\mbox{
\begin{tabular}{ccc}
\epsfxsize5cm
\epsffile[140 310 500 480]{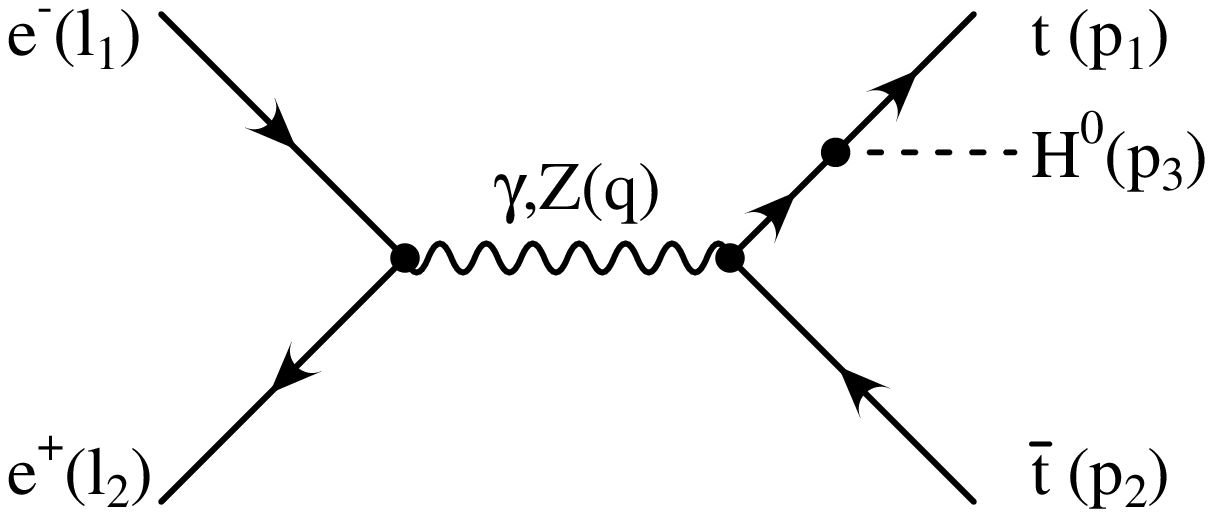}
\epsfxsize5cm
\epsffile[140 310 500 480]{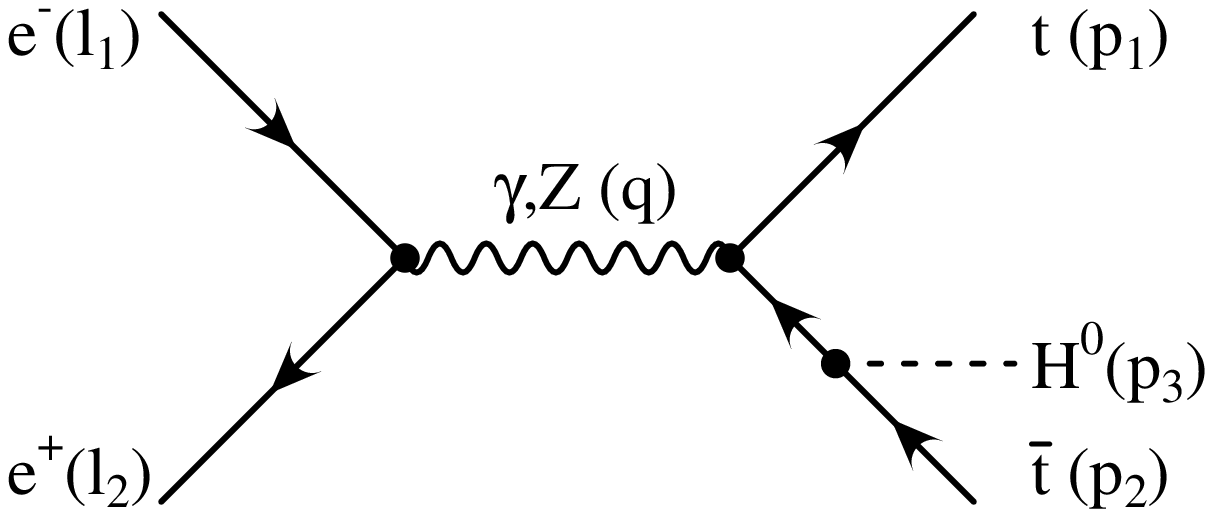}
\epsfxsize5cm
\epsffile[140 310 500 480]{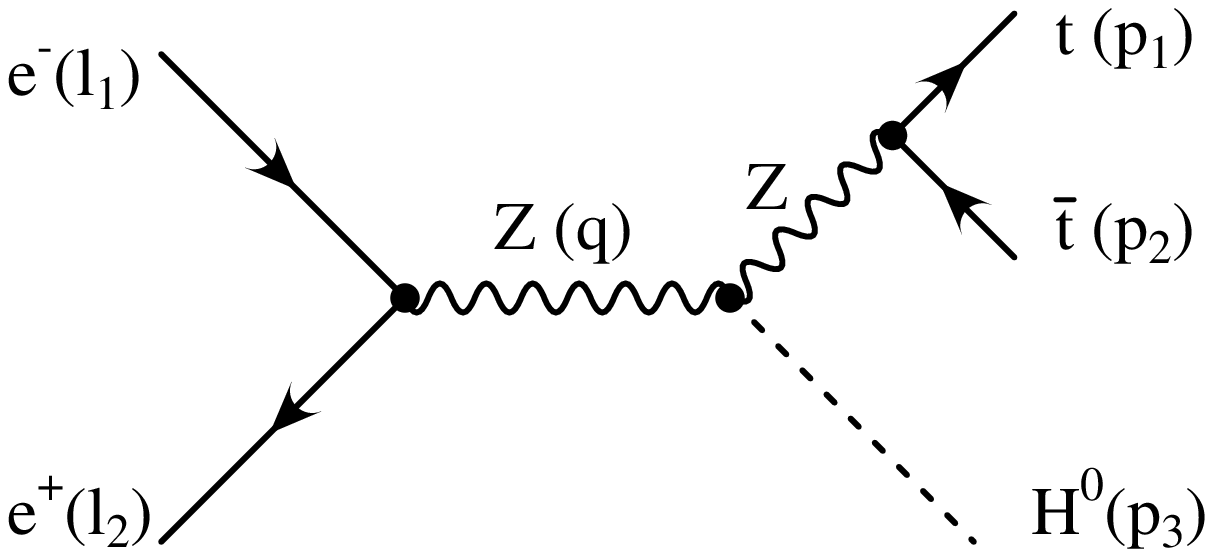}
\end{tabular}
}
\caption{\label{figh}\it Diagrams for \eettH.}
\end{center}
\end{figure}

\bigskip
\noindent%
For an accurate prediction of top pair production cross sections at a high
energy \Pep\Pem\ collider, various types of higher-order effects have
to be taken into account:
\begin{itemize}
\item the QCD corrections, which are treated perturbatively far from
      the threshold region, but require refinements on threshold and
      finite width effects close to the production threshold
      \cite{qcdcorr},
\item the electromagnetic bremsstrahlung corrections (QED corrections)
      with real and virtual photons inserted in the Born diagrams. They
      are complete at the 1-loop level in the virtual and soft photon
      part \cite{softphoton} as well as in the hard photon
      contribution \cite{hardphoton},
\item the genuine electroweak corrections, which consist of all
      electroweak 1-loop contributions to \eett, without virtual
      photons in the external charged fermion self energies and
      vertex corrections.
      These are also available as a complete set of 1-loop contribution
      \cite{softphoton,ew1,ew2}.
\end{itemize}
An important feature of the electroweak corrections is that they are
large and negative at high energies far above the \ttb\ threshold,
and thus lead to a sizeable reduction of the production cross-section.
At very high energies, on the other hand, there are also radiation
processes besides the conventional photon radiation which contribute
to inclusive top pair production \eettX:
The radiation of \PZ\ bosons (\reffi{figz}) and the radiation of
Higgs bosons (\reffi{figh}).

In previous work \cite{bremssearch,bremsyuk} these radiation processes
were studied with emphasis on searches for Higgs bosons and on
investigating the Yukawa interaction.
They may, however, also be considered as contributions to the
(inclusive) \ttb\ cross section at 1-loop order.
Since they constitute a positive contribution, they increase the
production rate of \ttb\ pairs according to $\Pep\Pem\to\Pt\Ptb(\PX)$
and hence compensate at least partially the negative terms from the
virtual corrections.

In this note  we present the
influence of the virtual electroweak 1-loop corrections on top pair
production and the contribution of $\Pep\Pem\to\Pt\Ptb\PZ,\Pt\Ptb\PH$
to the inclusive cross section and forward-backward asymmetry.
 In a first subsection the structure and
size of the virtual contributions are discussed.
Subsequently, we include the \PZ\ and \PH\ radiation processes and
show that the large and negative virtual contributions are
sizeably compensated by including the \PZ,\PH\ bremsstrahlung processes.

\bigskip
\noindent%
{\large \bf Virtual contributions}

\noindent%
The integrated cross section and the forward-backward asymmetry
for the process
$e^+(\pp)+e^-(\pe) \to t(\pt)+\bar{t}(\ptb)$
with purely electroweak virtual corrections
can be written in the form\footnote{Up to $O(\al^4)$.}
\bea
\sigma & = & \frac{4\pi\al(s)^2}{3s}\, N_C\,\beta\left\{
             \frac{1}{2}(3-\beta^2)\,\sigma_1(s)\, +\,
             \beta^2\,\sigma_2(s)\, +\,\Delta\sigma\right\} \nn \\
\AFB   & = & \frac{3}{4}\beta
             \frac{\sigma_3+\Delta\sigma'}{
             \frac{1}{2}(3-\beta^2)\,\sigma_1(s)\, +\,
             \beta^2\,\sigma_2(s)\, +\,\Delta\sigma}
\eea
where
\bea
s=(\pt+\ptb)^2, & & \nn \\
\beta=\sqrt{1-\frac{4\mt^2}{s}} & , &
\al(s)=\frac{\al}{1+\Pigr_{ferm}(s)}\equiv \frac{\al}{1-\Delta\al(s)}.
\eea
With $\Pigr_{ferm}=\Pigr-\Pigr_{bos}$ we denote the fermionic part
of the renormalized subtracted photon vacuum polarization.
$\sigma_1$, $\sigma_2$, and $\sigma_3$ contain the
lowest-order contribution and those one-loop corrections which can be
incorporated in effective photon-fermion and $Z$-fermion couplings,
i.e. the self-energies and the vertex corrections
to the V, A couplings.
The remaining terms, the top vertex corrections not
of V, A structure and the contribution from the $WW$ and $ZZ$
box diagrams, are collected in $\Delta\sigma$ and $\Delta\sigma'$
for the symmetric and the antisymmetric cross section.

The $\sigma_i$ can be written in the following way:
\bea
\sigma_1 & = & \left( Q_e^{V\,2} + Q_e^{A\,2} \right) \,
              Q_t^{V\,2} \nn \\
         &  & +\,2\,(Q_e^V V_e+Q_e^A A_e)\,Q_t^V V_t
              \, \frac{s}{s-\mz^2}   \nn \\
&  & +\left( V_e^2 +A_e^2\right)\, V_t^2\,
              \left(\frac{s}{s-\mz^2}\right)^2 \, ,\nn \\
 &  &  \nn \\
\sigma_2 & = & Q_e^{V\,2} \, Q_t^{A\,2} \nn \\
         &  & +\,2\,Q_e^V V_e\, Q_t^A A_t
              \, \frac{s}{s-\mz^2}   \nn \\
         &  & +\left( V_e^2 +A_e^2\right)\, A_t^2\,
              \left(\frac{s}{s-\mz^2}\right)^2\, , \nn \\
\sigma_3 & = & 2 Q_e^V Q_t^V A_e A_t \frac{s}{s-\mz^2}
              +4 V_e V_t A_e A_t \left(\frac{s}{s-\mz^2}\right)^2.
\eea

\noindent%
The effective coupling constants in these formulae are ($\Pf=\Pe,\Pt$)
\bea
Q_f^V & = & Q_f\left[1 -\frac{1}{2}\Pigr_{bos}(s) \right]
            -\fgf_V(s), \nn \\
Q_f^A & = &
            -\fgf_A(s), \nn \\
V_f & = & \bar{v}_f + \fzf_V(s), \nn\\
A_f & = & \bar{a}_f + \fzf_A(s)
\eea
with
\bea
\bar{a}_f & = & \left(\frac{\sqrt{2}\Gmu\mz}{4\pi\al(s)}\right)^{1/2}
          \rob^{1/2}\, I_3^f  \nn \\
\bar{v}_f & = & \left(\frac{\sqrt{2}\Gmu\mz}{4\pi\al(s)}\right)^{1/2}
          \rob^{1/2}\,(I_3^f-2 Q_f\bar{s}^2) \, .
\eea
They contain the propagator corrections
\bea
\rob & = & 1-\Delta r -\Pizr(s), \nn \\
\bar{s}^2 & = & s_W^2 -s_W c_W \,\Pigzr(s)
\eea
with the self-energies renormalized according to the on-shell scheme
\bea
\Pizr & = & \real\, \frac{\Szr(s)}{s-\mz} ,\nn \\
\Pigzr & = & \real\, \frac{\Sgzr(s)}{s}\, ,
\eea
together with the V, A form factors $F_{V,A}$ from the vertex corrections
(real part).
$\Delta r$ is the radiative correction to the Fermi coupling constant
in the on-shell scheme. The explicit expressions can be found in
\cite{softphoton,deltar}.

The Born cross section $\sigma^{(0)}$ can be obtained from the formulae
above by setting all the self-energies, vertex corrections, $\Delta\sigma$,
$\Delta\sigma'$ and $\Delta r$ in Eqs. (4-6) equal to zero.

\begin{figure}[ht]
\begin{center}
\mbox{
\epsffile{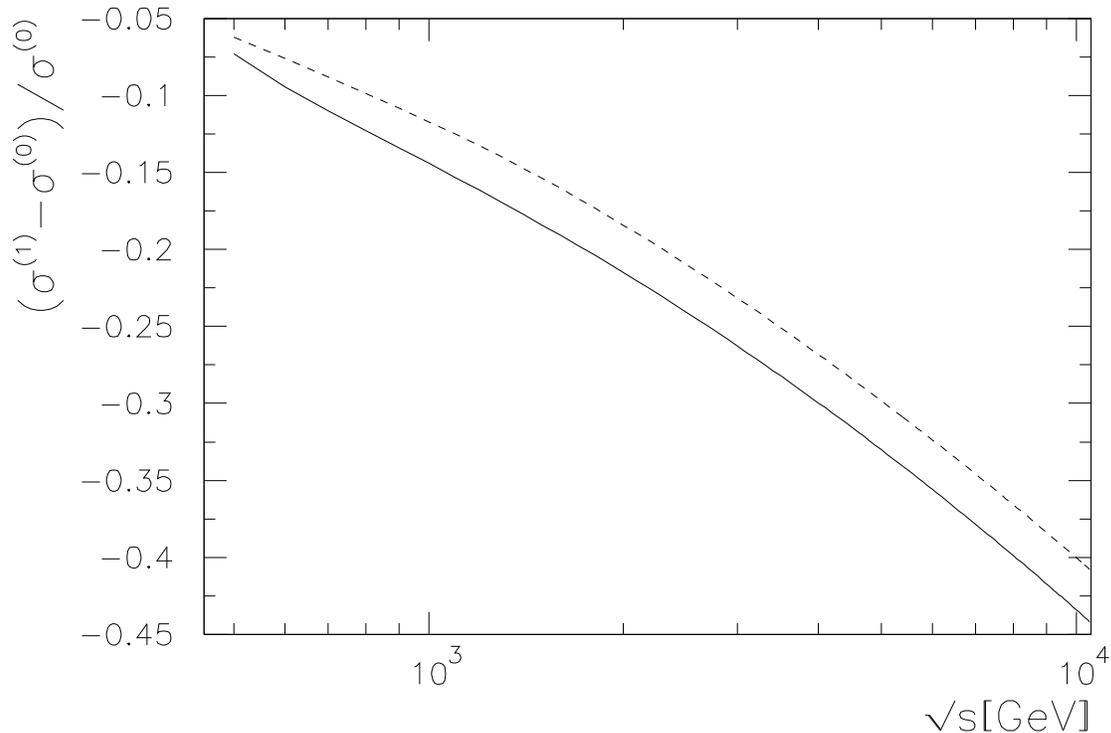}
}
\caption{\label{relcor}\it Relative electroweak corrections to \eett\
for $\mh=100\GeV$ (solid line) and $\mh=1000\GeV$ (dashed line).}
\end{center}
\end{figure}

The effect of the virtual electroweak corrections on the cross section
for \eett\ is displayed in \reffi{relcor}, where the relative deviation
from the "Born" cross section is shown.
Note that "Born" already includes the QED running of the effective
charge in the $\gamma$-exchange amplitude.
The results in \reffi{relcor} are thus the residual corrections which
are specific for the electroweak standard model.

\bigskip
\noindent%
{\large \bf Real contributions}

\noindent%
Next we consider the inclusive cross section for \ttb\ production in
association with Higgs and \PZ\ bremsstrahlung. We thereby restrict our
discussion to Higgs masses below $2\mt$, such that real Higgs
production with subsequent decay into \ttb\ cannot occur.
The latter case would be of specific interest for investigating the
Higgs Yukawa interaction, and was explicitly studied in
ref.~\cite{bremssearch}.

The bremsstrahlung processes are of higher order in the coupling constants
than the $2 \to 2$ process. The cross section at energies sufficiently
above the threshold reach 10\% and more of the lowest order cross section.
Their contribution to the \ttb\ final state hence becomes of the
same order as the virtual electroweak corrections.

For the computation of the cross sections corresponding to the amplitudes
in \reffis{figz} and \ref{figh} the following set of couplings has
been chosen:
\begin{eqnarray*}
\gamma\Pf\Pf & : & - \sqrt{4\pi\,\alpha(s)} \;\gamma^{\mu}Q_\Pf \\
\PZ\Pf\Pf    & : & - \sqrt{ \sqrt{2}\,\Gmu\,\mz^2}\; \gamma^{\mu}\;
                    [ (I_3 - 2 Q_f \, \sw^2 )-I_3\,\gamma_{5} ] \\
\PH\Pf\Pf    & : & - \sqrt{ \sqrt{2}\,\Gmu }\, \mf \\
\PH\PZ\PZ    & : & \sqrt{ 4\sqrt{2}\,\Gmu\,\mz^2 }\,\mz
\end{eqnarray*}
For $\sw^2$ the approximate expression
\beq
\sw^2=
\frac{1}{2}-\sqrt{ \frac{1}{4}-\frac{\pi\,\alpha(M_Z)}{\sqrt{2}\,\Gmu\,M_Z^2} }
=0.2311
\eeq
is used.

\begin{figure}[ht]
\begin{center}
\epsfxsize10cm
\mbox{\epsffile[30 210 540 570]{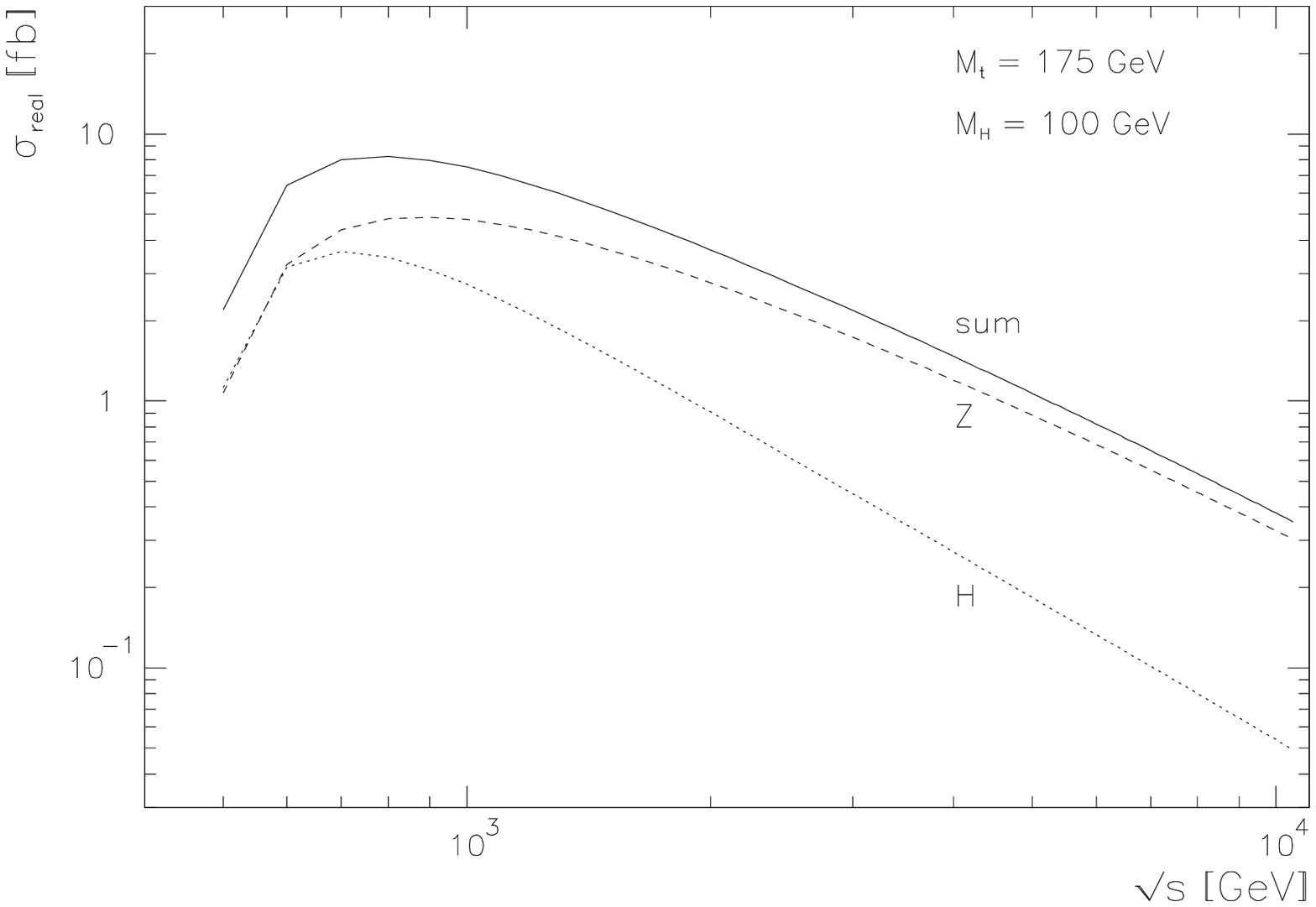}}
\caption{\label{realcross}\it Integrated cross section for \eettZ\ and \eettH.}
\end{center}
\end{figure}

In \reffi{realcross} we put together the integrated cross section for
\eettZ\ and \eettH\, as function of the energy $\sqrt{s}$.
The dominating contribution to the inclusive final state for
$\sqrt{s}\ge 1\TeV$ is from
the "\PZ-strahlung", as can be seen from \reffi{realcross}.

We now can define an inclusive cross section for \ttb\ production
in the following way:
\beq
\frac{d\sigma}{d\Omega}(\ttb)=\frac{d\sigma_V}{d\Omega}(\ttb)
      +\int d^3 p_3 \frac{d\sigma(\ttb\PH)}{d\Omega d^3 p_3}
      +\int d^3 p_3 \frac{d\sigma(\ttb\PZ)}{d\Omega d^3 p_3}
\eeq
where $d\Omega=d\cos\theta_t\;d\phi$ is the solid angle of the outgoing top,
and $\theta_t$ the scattering angle between $e^-$ and $t$.
$d\sigma_V$ denotes the 2-particle final states including the
virtual contributions, and $d^3 p_3$ is the phase space element for
\PH\ and \PZ\,, respectively.

\noindent%
The integrated cross section is obtained as
\beq
\sigma_{\ttb}=\sigma_V(\ttb)+\sigma(\eettH)+\sigma(\eettZ).
\eeq

\noindent%
The forward-backward asymmetry \AFB\ is given by
\beq
A_{FB}\;=\; \frac{1}{\sigma_{\ttb}}
\left(  \int_{0}^{1}  d\cos\theta_t \; \frac{d\sigma(\ttb)}{d\cos\theta_t}
      - \int_{-1}^{0} d\cos\theta_t \; \frac{d\sigma(\ttb)}{d\cos\theta_t}
 \right) 
\eeq
with
\beq
  \frac{d\sigma(\ttb)}{d\cos\theta_t} \;=\;
 \int_{0}^{2\pi} d\phi \; \frac{d\sigma}{d\Omega}(\ttb)
\eeq

\begin{figure}[ht]
\begin{center}
\mbox{\epsffile{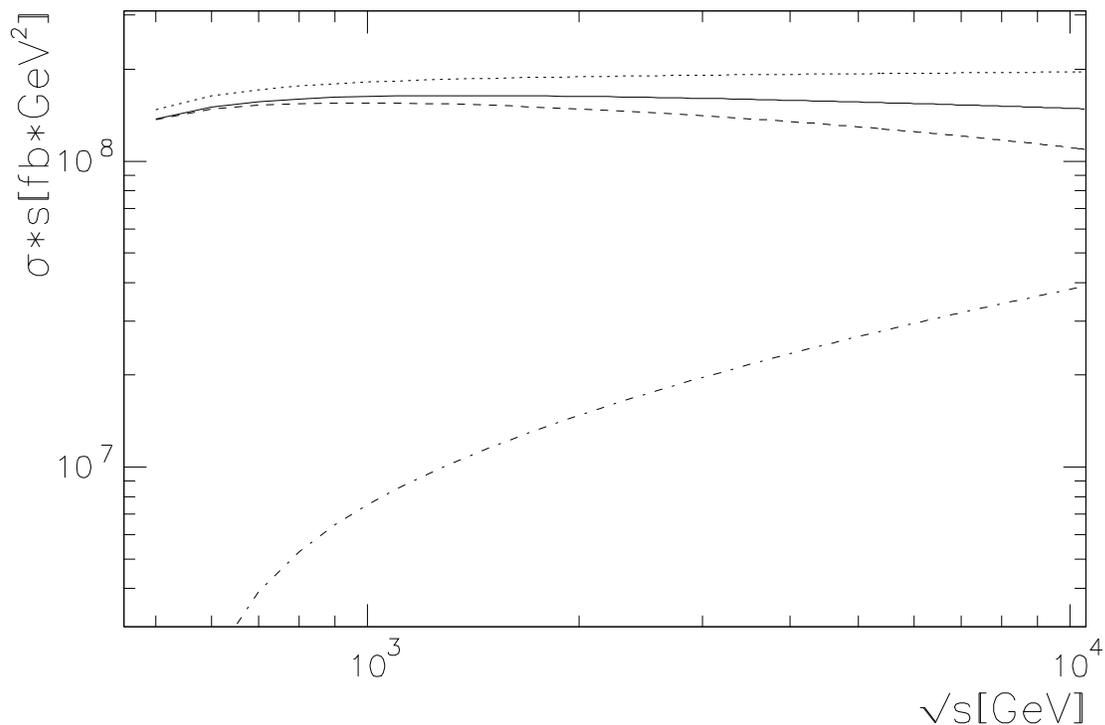}}
\caption{\label{sigtot}\it Integrated cross section (times $s$).
The dotted line shows the Born cross section $\sigma^{(0)}$,
the dashed line contains all electroweak 1-loop virtual corrections,
the dash-dotted line depicts the real contributions from \eettH\
and \eettZ\,, and the solid line shows the inclusive cross section $\sigma$.}
\end{center}
\end{figure}

\begin{figure}[ht]
\begin{center}
\mbox{\epsffile{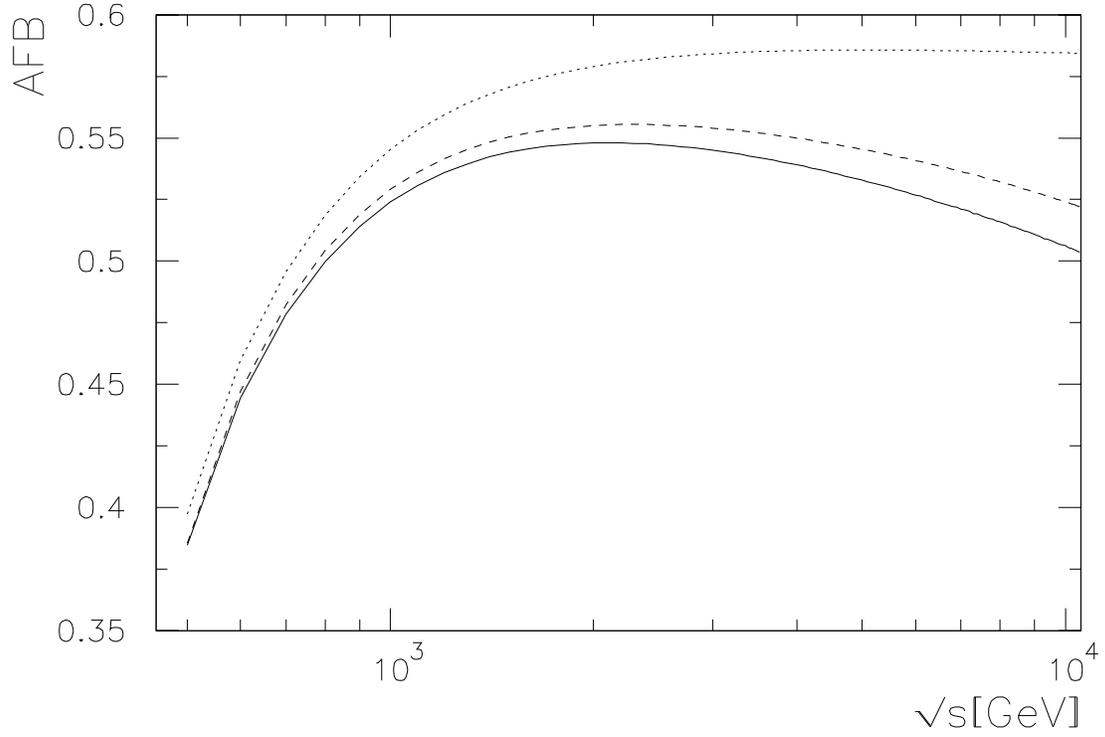}}
\caption{\label{afbtot}\it Forward-backward asymmetry.
The dotted line show the Born prediction, the dashed line contains
also all electroweak 1-loop virtual corrections, and the solid
line depicts the result for the inclusive production.}
\end{center}
\end{figure}

The results are displayed in \reffis{sigtot}, \ref{afbtot}.
As one can see from \reffi{sigtot}, the virtual and real
contributions to the integrated cross section cancel each other
to a large extent.
For \AFB, however, the situation is different. \reffi{afbtot}
contains \AFB\ in Born order, with virtual corrections, and after including
the real contributions. For $A_{FB}$, the totally inclusive \ttb\
asymmetry deviates more from the Born result than the one
with only virtual corrections to \eett.

\mbox{}

Thereby, \AFB\ in Born approximation is obtained from Eq.~(1) with
expressions (3-6), where all self energies, vertex corrections,
$\Delta r$ and $\Delta\sigma$,$\Delta\sigma'$ are put to zero.


\clearpage

\begin{thebibliography}{9}
\frenchspacing
\newcommand{\zp}[3]{{\sl Z.~Phys.} {\bf C#1} (19#2) #3}
\newcommand{\np}[3]{{\sl Nucl.~Phys.} {\bf B#1} (19#2) #3}
\newcommand{\pl}[3]{{\sl Phys.~Lett.} {\bf B#1} (19#2) #3}
\newcommand{\pr}[3]{{\sl Phys.~Rev.} {\bf D#1} (19#2) #3}
\newcommand{\fp}[3]{{\sl Fortschr.~Phys.} {\bf #1} (19#2) #3}
\newcommand{\cpc}[3]{{\sl Comput.~Phys.~Commun.} {\bf #1} (19#2) #3}
\newcommand{\rep}[3]{{\sl Phys.~Rep.} {\bf #1} (19#2) #3}
\newcommand{\pis}[3]{{\sl Pi'sma v Zh.~Eksp.~Teor.~Fiz.} {\bf #1} (19#2) #3}
\newcommand{\ya}[3]{{\sl Yad.~Fiz.} {\bf #1} (19#2) #3}
 
\bibitem{qcdcorr} 
J.~Jersak, E.~Laerman, P.M.~Zerwas, \pr{25}{80}{1218}; \\
S.~G\"usken, J.H.~K\"uhn, P.M.~Zerwas, \pl{155}{85}{185}; \\
J.H.~K\"uhn, P.M.~Zerwas, \rep{167}{88}{321}; \\
V.S.~Fadin, V.A.~Khoze, \pis{46}{87}{417}; \ya{48}{88}{487}; \\
V.S.~Fadin, V.A.~Khoze, T.~Sj\"ostrand, \zp{48}{90}{613}; \\
V.S.~Fadin, O.I.~Yakovlev, Novosibirsk preprint IYF 90-138 (1990); \\
W.~Kwong, \pr{43}{91}{1488}; \\
H.~Inazawa,, T.~Morii, J.~Morishita, \pl{203}{88}{279}; \zp{42}{89}{569}; \\
K.~Hagiwara et al., \np{344}{90}{1}; \\
J.~Feigenbaum, \pr{43}{91}{264}; \\
M.J.~Strassler, M.E.~Peskin, \pr{43}{91}{1500}; \\
J.H. K\"uhn, Proceedings of the {\it Workshop on on Physics and
Experiments with Linear $e^+e^-$ Colliders}, 
Waikoloa, Hawaii 1993, eds.: F.A. Harris, S.L. Olsen, S. Pakvasa,
X. Tata
\bibitem{softphoton}
W.~Beenakker, W.~Hollik and S.C.~van der Marck, \np{365}{91}{24}.
\bibitem{hardphoton}
A.A.~Akhundov, D.Y.~Bardin and A.~Leike, \pl{261}{91}{321}; \\
A. Arbuzov, D. Bardin, A. Leike, {\it Mod.\ Phys.\ Lett.\ }
{\bf A7} (1992) 2029; E: {\bf A9} (1994) 1515.  
\bibitem{ew1}
W.~Beenakker and W.~Hollik, \pl{269}{91}{425}.
\bibitem{ew2}
W.~Beenakker, A.~Denner and A.~Kraft, \np{410}{93}{219}.
\bibitem{deltar}
W.~Hollik, \fp{38}{90}{165}; \\
M.~Consoli, W.~Hollik, F.~Jegerlehner, in: \PZ\ Physics at LEP1,
CERN 89-08(1989), eds.~G.~Altarelli, R.~Kleiss, C.~Verzegnassi.
\bibitem{bremssearch}
K.~Hagiwara, H.~ Murayama, I.~ Watanabe, \np{367}{91}{257}.
\bibitem{bremsyuk}
A.~Djouadi, J.~Kalinowski, P.M.~Zerwas, \zp{54}{92}{255}
\end{thebibliography}
\end{document}